\documentclass{jkas}

\def\year{2014} % publication year
\def\month{December} % publication month
\def\volume{47} % journal volume
\def\issue{6} % journal issue
\def\beginpage{293} % first page of article
\def\endpage{302} % last page of article
\def\received{November 20, 2014} % date paper was received by JKAS
\def\accepted{November 28, 2014} % date of acceptance
\date{Received \received; accepted \accepted}

\newcommand{\m}{\mbox{\hspace{2.0mm}}}

\newcommand{\xx}{\mbox{\hspace{3.5mm}}}
\newcommand{\e}{\mbox{\hspace{4.0mm}}}
\newcommand{\ee}{\mbox{\hspace{4.5mm}}}

\newcommand{\nn}{\mbox{\hspace{5.5mm}}}

\usepackage{flushend}

\title{First Detection of 22 GHz H$_{2}$O Masers in TX Camelopardalis}

\author[]{Se-Hyung Cho}
\author[]{Jaeheon Kim}
\author[]{Youngjoo Yun}

\affil[]{Korean VLBI Network, Korea Astronomy and Space Science Institute, 776 Daedukdae-ro, Yuseong-gu, Daejeon 305-348, Korea; \email{cho@kasi.re.kr, jhkim@kasi.re.kr, yjyun@kasi.re.kr}}

\def\corrauthor{S.-H. Cho}

\def\runningauthor{Cho, Kim \& Yun}

\def\runningtitle{First Detection of 22 GHz H$_{2}$O Masers in TX Cam}

\def\keywords{circumstellar matter --- masers --- radio lines --- stars: late-type}

\def\abstracttext{
Simultaneous time monitoring observations of H$_{2}$O $6_{16}-5_{23}$, SiO $J$ = 1--0, 2--1, 3--2, and $^{29}$SiO $v$ = 0, $J$ = 1--0 lines were carried out in the direction of the Mira variable star 
TX Cam with the Korean VLBI Network single dish radio telescopes. For the first time, the H$_{2}$O maser emission from TX Cam was detected near the stellar velocity at five epochs from April 10, 2013 
($\phi$ = 3.13) to June 4, 2014 ($\phi$ = 3.89) including minimum optical phases. The intensities of H$_{2}$O masers are very weak compared to SiO masers. The variation of peak antenna temperature ratios 
among SiO $v$ = 1, $J$ = 1--0, $J$ = 2--1, and $J$ = 3--2 masers is investigated according to their phases. The shift of peak velocities of H$_{2}$O and SiO masers with respect to the stellar velocity is also 
investigated according to observed optical phases. The H$_{2}$O maser emission occurs around the stellar velocity during our monitoring interval. On the other hand, the peak velocities of SiO masers show 
a spread compared to the stellar velocity. The peak velocities of SiO $J$ = 2--1, and $J$ = 3--2 masers show a smaller spread with respect to the stellar velocity than those of SiO $J$ = 1--0 masers. These 
simultaneous observations of multi-frequencies will provide a good constraint for maser pumping models and a good probe for investigating the stellar atmosphere and envelope according to their different 
excitation conditions.
}

\begin{document}
\jkashead %% set title, authors, abstract, etc.

\section{Introduction\label{sec:intro}}

TX Cam is a cool Mira variable star with a late spectral type M8$-$M10. It showed an optical variation from V$_{max}$ = 11.60 to V$_{min}$ =16.20 in magnitude with a period of 557.4 d (Kholopov et al. 1985). 
Its distance was estimated to be 390 pc from the period-luminosity relation (Olivier et al. 2001). The mass-loss rate was obtained to be $\sim$ 1.1 $\times$ 10$^{-6}$ $M_{\odot}$ yr$^{-1}$ by Knapp \& Morris (1985). 

Silicon monoxide (SiO) maser emission from TX Cam was detected by Spencer et al. (1977) for the first time. Subsequently, various vibrational and rotational transitions of $^{28}$SiO, $^{29}$SiO, and 
$^{30}$SiO masers including the rare masers of $^{28}$SiO $v$ = 2, $J$ = 2--1 and $^{29}$SiO $v$ = 1, $J$ = 1--0 lines were detected from this star (Jewell et al. 1987, Herpin et al. 1998, Gray et al. 1999, 
Cho \& Ukita 1995, 1998, Cho et al. 1998, Cho \& Kim 2012). Single dish monitoring of SiO masers toward evolved stars including this star has been performed in the SiO $v$ = 1, 2, $J$ = 1--0 lines by Alcolea et al. (1999) 
and  Pardo et al. (2004). Kang et al. (2006) also performed monitoring observations in both SiO $v$ = 1, 2, 3, $J$ = 2--1 and $J$ = 3--2 masers using the TRAO 14 m telescope. Alcolea et al. (1999) and  
Pardo et al. (2004) showed that the SiO intensity maxima are in agreement with those of NIR maxima with phase lags of 0.05 -- 0.20 supporting the radiative pumping of SiO masers against collisional pumping. 
Kang et al. (2006) reported that the time variability of $v$ = 1, $J$ = 2--1 and $J$ = 3--2 masers is very different, i. e., main peak components of the $v$ = 1, $J$ = 2--1 maser show dominant red-shifted features 
compared to the $v$ = 1, $J$ = 3--2 maser. For VLBI monitoring observations, a movie of the SiO ($v$ = 1, $J$ = 1--0) masers around TX Cam using the VLBA was presented by Diamond \& Kemball (2003) and directly 
showed stellar pulsation motions traced by SiO masers. Gonidakis et al. (2010, 2013) also presented new and final versions of the SiO maser movies exhibiting the existence of shock waves and their effect on the morphology 
and kinematics of the gas. However, the H$_{2}$O maser (Bowers \& Hagen 1984, Benson \& Little-Marenin 1996, Shintani et al. 2008) as well as the OH maser (Olnon et al. 1980) has not been detected in TX Cam. This is not the case of a large number of oxygen-rich AGB stars in which both H$_{2}$O and OH masers are detected together with SiO masers. Benson \& Little-Marenin (1996) did not detected the H$_{2}$O maser despite repeated attempts. They 
considered this missed detection quite strange, despite a high mass-loss rate, a relatively close distance, and a well-developed shell. Therefore, we included TX Cam in our simultaneous monitoring programs of 22 GHz H$_{2}$O 
and 43, 86, 129 GHz SiO masers using the Korean VLBI Network (KVN) single dish telescopes in order to confirm the existence of the 22 GHz H$_{2}$O maser and the variation characteristics of the maser properties among 
SiO $J$ = 1--0, $J$ = 2--1, and $J$ = 3--2 masers through systematic monitoring observations.

In this paper, we present the first detection of the 22 GHz H$_{2}$O $6_{16}-5_{23}$ maser emission and results of simultaneous observations of $^{28}$SiO $v$ = 0, 1, 2, $J$ = 1--0, $J$ = 2--1, $J$ = 3--2, $^{29}$SiO $v$ = 0, 
$J$ = 1--0 and H$_{2}$O $6_{16}-5_{23}$ masers in the direction of TX Cam. In Section 2, we describe the observations. In Section 3, observational results are presented. In Section 4, we discuss the detection of H$_{2}$O maser 
emission, the intensity ratio and the peak velocity pattern among SiO masers according to the optical phases. A summary is given in Section 5.

\section{Observations\label{sec:obs}}

Simultaneous monitoring observations of H$_{2}$O and SiO masers have been performed from June 2009 to June 2014. The SiO $J$ = 2--1 and $J$ = 3--2 lines began to be observed from May 27, 2012 after installation of 
86/129 GHz band receivers. Therefore, simultaneous monitoring of H$_{2}$O $6_{16}-5_{23}$ and $^{28}$SiO $v$ = 1, 2, $J$ = 1--0, $^{29}$SiO $v$ = 0, $J$ = 1--0 maser lines were performed from June 10, 2009 to April 13, 2012 
and simultaneous monitoring of H$_{2}$O $6_{16}-5_{23}$ and $^{28}$SiO $v$ = 1, 2, $J$ = 1--0; $v$ = 1, $J$ = 2--1, $J$ = 3--2 maser lines performed from May 27, 2012 to June 4, 2014 except the SiO $J$ = 3--2 maser 
 which was observed in 2012 July and October. Simultaneous observations of SiO $v$ = 0, $J$ = 1--0; $v$ = 2, $J$ = 2--1, $J$ = 3--2 lines were added in May 2012 and SiO $v$ = 0, $J$ = 1--0, $J$ = 2--1 lines added in July 2012. In total, 15 epoch 
data were obtained. 

We used one of the KVN Yonsei, Ulsan, and Tamna 21 m single dish telescopes based on our observational dates. The KVN antenna optics were designed for supporting simultaneous observations in four bands of 22, 43, 86, and 
129 GHz using three quasi-optical low-pass filters (Han et al. 2008, 2013). The low pass filter 1 (LPF1) reflects the beam for the 86/129 GHz bands and the beam for the 22 GHz and 43 GHz bands. Likewise, LPF2 
and LPF3 split beams into 22/43 GHz and 86/129 GHz, respectively. The half power beam widths and aperture efficiencies of the three KVN telescopes at four bands were given in the KVN home page (\url{http://kvn-web.kasi.re.kr/}). 
We adopted an average values of the three telescopes because their parameters are similar. Hence, the average half power beam widths and aperture efficiencies are 123$''$, 0.58 (at 22 GHz) and 62$''$, 0.61 (at 43 GHz), 
32$''$, 0.50 (at 86 GHz) and 23$''$, 0.35 (at 129 GHz), respectively. The pointing accuracy was checked every 2 hours using the source itself. We used the cryogenic 22, 43, 86 GHz High Electron Mobility Transistor (HEMT) 
receivers and the Superconductor-Insulator-Superconductor (SIS) 129 GHz receiver with both right and left circular polarized feeds (Han et al. 2013). Only the left circular polarized feed was used during our observations. 

The SSB system noise temperatures of a single side-band range from 72 K to 156 K (at 22 GHz), from 96 K to 235 K (at 43 GHz), from 175 K to 385 K (at 86 GHz), and from 189 K to 566 K (at 129 GHz) depending on 
weather conditions and elevations. We used a digital spectrometer with total band widths chosen from one 64 MHz mode for the H$_{2}$O $6_{16}-5_{23}$ line and three 64 MHz modes for the SiO $v$ = 1, 2, and $^{29}$SiO 
$v$ = 0, $J$ = 1--0 lines until the April 2012 observation with only available 22 and 43 GHz modes. After that period, for the four bands (22, 43, 86, and 129 GHz), we used three 32 MHz modes for the H$_{2}$O $6_{16}-5_{23}$, 
SiO $v$ = 1, 2, $J$ = 1--0 lines and two 64 MHz modes for the SiO $v$ = 1, $J$ = 2--1 and $J$ = 3--2 lines, respectively. These band widths correspond to the radial velocity ranges of 860 km s$^{-1}$ (at 22 GHz), 
440 km s$^{-1}$ (at 43 GHz). The velocity resolutions of each band are 0.21 km s$^{-1}$ (at 22 GHz), 0.11 km s$^{-1}$ (at 43 GHz) until the April 2012 observation mode. After that, for the four band modes, the radial 
velocity ranges are 440 km s$^{-1}$ (at 22 GHz), 222 km s$^{-1}$ (at 43 GHz), 222 km s$^{-1}$ (at 86 GHz), and 148 km s$^{-1}$ (at 129 GHz) with velocity resolutions of 0.11 km s$^{-1}$ (at 22 GHz), 0.05 km s$^{-1}$ 
(at 43 GHz), 0.05 km s$^{-1}$ (at 86 GHz), and 0.036 km s$^{-1}$ (at 129 GHz), respectively. All the spectra were Hanning-smoothed for velocity resolutions of 0.44 $-$ 0.57 km s$^{-1}$.

The chopper wheel method was used for the data calibration; it corrects the atmospheric attenuation and the antenna gain variations, depending on elevation in order to yield an antenna temperature $T^\ast_A$. The integration time 
was 60$-$120 minutes to achieve the sensitivity of $\sim$0.06 K at the 3$\sigma$ level. The average conversion factors of the three telescopes from the antenna temperature to the flux density are about 13.8 Jy K$^{-1}$ 
at 22 GHz, 13.1 Jy K$^{-1}$ at 43 GHz, 15.9 Jy K$^{-1}$ at 86 GHz, and 22.8 Jy K$^{-1}$ at 129 GHz, respectively.  

%%% TABLE 1 %%%%%%%%%%%%%%%%%%%%%%%%%%%%%%%%%%%%%%%%%%%%%%%%%%%%%%%%%%%%%%%%%%%%%%%%%%%%%%%%%%%%%%%%%%%%

\begin{table}[t!]
\caption{H$_{2}$O and SiO transitions and rest frequencies used for the observations\label{tab:jkastable1}}
\centering
\begin{tabular}{ccr}
\toprule
Molecule & Transition & Frequency (GHz) \\
\midrule

H$_{2}$O   & 6$_{1,6}$--5$_{2,3}$ &  22.235080\e \\
$^{28}$SiO & $v$ = 0, $J$ = 1--0  &  43.423858\e \\
           & $v$ = 1, $J$ = 1--0  &  43.122080\e \\
           & $v$ = 2, $J$ = 1--0  &  42.820587\e \\
           & $v$ = 0, $J$ = 2--1  &  86.846998\e \\
           & $v$ = 1, $J$ = 2--1  &  86.243442\e \\
           & $v$ = 2, $J$ = 2--1  &  85.640452\e \\
           & $v$ = 1, $J$ = 3--2  & 129.363359\e \\
           & $v$ = 2, $J$ = 3--2  & 128.458891\e \\
$^{29}$SiO & $v$ = 0, $J$ = 1--0  &  42.879916\e \\

\bottomrule
\end{tabular}
\end{table}

%%%%%%%%%%%%%%%%%%%%%%%%%%%%%%%%%%%%%%%%%%%%%%%%%%%%%%%%%%%%%%%%%%%%%%%%%%%%%%%%%%%%%%%%%%%%%%%%%%%%%%%%%

\section{Observational Results\label{sec:result}}

%%% TABLE 2 %%%%%%%%%%%%%%%%%%%%%%%%%%%%%%%%%%%%%%%%%%%%%%%%%%%%%%%%%%%%%%%%%%%%%%%%%%%%%%%%%%%%%%%%%%%%%

\begin{table*}[tp]
\caption{Results of the H$_{2}$O and SiO maser monitoring observations\label{tab:jkastable2}}
\centering
\begin{tabular}{crcrrrc}
\toprule
Molecule and Transition & $T_{\rm A}^{\ast}$(peak) & rms & $\int\! T_{\rm A}^{\ast}dv$\e & $V_{\rm LSR}$(peak) & $V_{\rm LSR}$(peak)$-V_{\ast}$ & Date(phase) \\
                        & (K)\e                    & (K) & (K km s$^{-1}$)               & (km s$^{-1}$)\m     & (km s$^{-1}$)\ee               & (yymmdd)    \\
         (1)            & (2)\ee                   & (3) & (4)\nn                        & (5)\nn              & (6)\nn\e                       & (7)         \\
\midrule

H$_{2}$O 6$_{1,6}$--5$_{2,3}$  & $\cdots$\ee & 0.08 & $\cdots$\nn & $\cdots$\nn & $\cdots$\nn\e & 090610(0.62) \\
                               & $\cdots$\ee & 0.02 & $\cdots$\nn & $\cdots$\nn & $\cdots$\nn\e & 101211(1.60) \\
                               & $\cdots$\ee & 0.05 & $\cdots$\nn & $\cdots$\nn & $\cdots$\nn\e & 110222(1.74) \\
                               & $\cdots$\ee & 0.02 & $\cdots$\nn & $\cdots$\nn & $\cdots$\nn\e & 111018(2.16) \\
                               & $\cdots$\ee & 0.01 & $\cdots$\nn & $\cdots$\nn & $\cdots$\nn\e & 120102(2.30) \\
                               & $\cdots$\ee & 0.02 & $\cdots$\nn & $\cdots$\nn & $\cdots$\nn\e & 120413(2.48) \\
                               & $\cdots$\ee & 0.03 & $\cdots$\nn & $\cdots$\nn & $\cdots$\nn\e & 120527(2.56) \\
                               & $\cdots$\ee & 0.04 & $\cdots$\nn & $\cdots$\nn & $\cdots$\nn\e & 120704(2.63) \\
                               & $\cdots$\ee & 0.01 & $\cdots$\nn & $\cdots$\nn & $\cdots$\nn\e & 121210(2.91) \\
                               &  0.12\xx    & 0.02 &  0.23\ee    & 11.2\ee     &    0.2\nn\xx  & 130410(3.13) \\
                               &  0.07\xx    & 0.02 &  0.08\ee    & 10.8\ee     & $-$0.2\nn\xx  & 130915(3.42) \\
                               &  0.06\xx    & 0.01 &  0.11\ee    & 12.5\ee     &    1.5\nn\xx  & 131125(3.54) \\
                               & $\cdots$\ee & 0.01 & $\cdots$\nn & $\cdots$\nn & $\cdots$\nn\e & 140211(3.68) \\
                               &  0.09\xx    & 0.02 &  0.22\ee    & 11.2\ee     &    0.2\nn\xx  & 140407(3.78) \\
                               &  0.15\xx    & 0.03 &  0.35\ee    & 11.3\ee     &    0.3\nn\xx  & 140604(3.89) \\ \addlinespace
$^{28}$SiO $v$ = 0, $J$ = 1--0 &  0.26\xx    & 0.04 &  2.55\ee    & 12.9\ee     &    1.9\nn\xx  & 120527(2.56) \\
                               &  0.20\xx    & 0.02 &  4.07\ee    & 10.7\ee     & $-$0.3\nn\xx  & 120704(2.63) \\ \addlinespace
$^{28}$SiO $v$ = 1, $J$ = 1--0 &  3.53\xx    & 0.08 &  9.44\ee    &  6.6\ee     & $-$4.4\nn\xx  & 090610(0.62) \\
                               &  5.57\xx    & 0.03 & 15.42\ee    & 11.0\ee     &    0.0\nn\xx  & 101211(1.60) \\
                               &  9.10\xx    & 0.05 & 29.18\ee    & 11.1\ee     &    0.1\nn\xx  & 110222(1.74) \\
                               & 10.85\xx    & 0.02 & 72.17\ee    & 11.6\ee     &    0.6\nn\xx  & 111018(2.16) \\
                               &  7.26\xx    & 0.02 & 47.71\ee    &  7.7\ee     & $-$3.3\nn\xx  & 120102(2.30) \\
                               &  0.88\xx    & 0.03 &  5.14\ee    &  6.5\ee     & $-$4.5\nn\xx  & 120413(2.48) \\
                               &  2.28\xx    & 0.02 & 13.14\ee    &  7.2\ee     & $-$3.8\nn\xx  & 120527(2.56) \\
                               &  2.49\xx    & 0.03 & 15.29\ee    &  5.1\ee     & $-$5.9\nn\xx  & 120704(2.63) \\
                               &  6.08\xx    & 0.04 & 47.31\ee    & 10.4\ee     & $-$0.6\nn\xx  & 121210(2.91) \\
                               & 26.64\xx    & 0.01 & 95.93\ee    & 12.5\ee     &    1.5\nn\xx  & 130410(3.13) \\
                               & 14.53\xx    & 0.01 & 45.85\ee    & 12.5\ee     &    1.5\nn\xx  & 130915(3.42) \\
                               & 11.45\xx    & 0.01 & 38.01\ee    & 13.0\ee     &    2.0\nn\xx  & 131125(3.54) \\
                               & 15.91\xx    & 0.01 & 51.59\ee    & 13.4\ee     &    2.4\nn\xx  & 140211(3.68) \\
                               & 19.98\xx    & 0.01 & 66.76\ee    & 12.5\ee     &    1.5\nn\xx  & 140407(3.78) \\
                               & 19.48\xx    & 0.02 & 58.20\ee    & 12.5\ee     &    1.5\nn\xx  & 140604(3.89) \\ \addlinespace
$^{28}$SiO $v$ = 2, $J$ = 1--0 &  2.78\xx    & 0.09 & 16.61\ee    & 10.6\ee     & $-$0.4\nn\xx  & 090610(0.62) \\
                               &  8.62\xx    & 0.03 & 20.18\ee    & 10.9\ee     & $-$0.1\nn\xx  & 101211(1.60) \\
                               &  9.19\xx    & 0.05 & 24.45\ee    & 10.6\ee     & $-$0.4\nn\xx  & 110222(1.74) \\
                               & 16.15\xx    & 0.02 & 77.43\ee    & 11.2\ee     &    0.2\nn\xx  & 111018(2.16) \\
                               &  9.89\xx    & 0.02 & 47.41\ee    &  7.2\ee     & $-$3.8\nn\xx  & 120102(2.30) \\
                               &  3.25\xx    & 0.03 & 12.18\ee    &  7.6\ee     & $-$3.4\nn\xx  & 120413(2.48) \\
                               &  1.70\xx    & 0.04 &  7.99\ee    &  5.1\ee     & $-$5.9\nn\xx  & 120527(2.56) \\
                               &  2.35\xx    & 0.02 &  9.51\ee    &  5.1\ee     & $-$5.9\nn\xx  & 120704(2.63) \\
                               &  9.34\xx    & 0.04 & 38.93\ee    & 10.7\ee     & $-$0.3\nn\xx  & 121210(2.91) \\
                               & 21.98\xx    & 0.01 & 89.51\ee    & 12.5\ee     &    1.5\nn\xx  & 130410(3.13) \\
                               & 14.60\xx    & 0.01 & 50.66\ee    & 12.9\ee     &    1.9\nn\xx  & 130915(3.42) \\
                               & 18.02\xx    & 0.01 & 49.88\ee    & 13.0\ee     &    2.0\nn\xx  & 131125(3.54) \\
                               & 17.76\xx    & 0.01 & 45.35\ee    & 13.0\ee     &    2.0\nn\xx  & 140211(3.68) \\
                               & 18.12\xx    & 0.01 & 45.39\ee    & 12.9\ee     &    1.9\nn\xx  & 140407(3.78) \\
                               & 22.32\xx    & 0.03 & 51.44\ee    & 13.0\ee     &    2.0\nn\xx  & 140604(3.89) \\ \addlinespace
$^{28}$SiO $v$ = 0, $J$ = 2--1 &  0.41\xx    & 0.03 & 10.52\ee    & 10.3\ee     & $-$0.7\nn\xx  & 120704(2.63) \\ \addlinespace
$^{28}$SiO $v$ = 1, $J$ = 2--1 &  8.13\xx    & 0.03 & 23.90\ee    & 10.4\ee     & $-$0.6\nn\xx  & 120527(2.56) \\
                               &  5.40\xx    & 0.03 & 18.38\ee    & 10.8\ee     & $-$0.2\nn\xx  & 120704(2.63) \\
                               &  9.21\xx    & 0.01 & 43.54\ee    &  9.5\ee     & $-$1.5\nn\xx  & 121210(2.91) \\
                               & 11.43\xx    & 0.02 & 55.91\ee    & 11.5\ee     &    0.5\nn\xx  & 130410(3.13) \\
                               &  8.59\xx    & 0.02 & 38.92\ee    & 10.4\ee     & $-$0.6\nn\xx  & 130915(3.42) \\
                               &  6.86\xx    & 0.01 & 34.61\ee    &  9.9\ee     & $-$1.1\nn\xx  & 131125(3.54) \\
                               &  3.66\xx    & 0.02 & 24.51\ee    & 12.0\ee     &    1.0\nn\xx  & 140211(3.68) \\
                               &  6.13\xx    & 0.02 & 32.47\ee    & 11.6\ee     &    0.6\nn\xx  & 140407(3.78) \\

\bottomrule
\end{tabular}
\end{table*}

\setcounter{table}{1}
\begin{table*}[t!]
\caption{(Continued)\label{tab:jkastable2-cont}}
\centering
\begin{tabular}{crcrrrc}
\toprule
Molecule and Transition & $T_{\rm A}^{\ast}$(peak) & rms & $\int\! T_{\rm A}^{\ast}dv$\e & $V_{\rm LSR}$(peak) & $V_{\rm LSR}$(peak)$-V_{\ast}$ & Date(phase) \\
                        & (K)\e                    & (K) & (K km s$^{-1}$)               & (km s$^{-1}$)\m     & (km s$^{-1}$)\ee               & (yymmdd)    \\
         (1)            & (2)\ee                   & (3) & (4)\nn                        & (5)\nn              & (6)\nn\e                       & (7)         \\
\midrule

                               &  9.97\xx    & 0.03 & 38.56\ee    & 11.6\ee     &    0.6\nn\xx  & 140604(3.89) \\ \addlinespace
$^{28}$SiO $v$ = 2, $J$ = 2--1 & $\cdots$\ee & 0.05 & $\cdots$\nn & $\cdots$\nn & $\cdots$\nn\e & 120527(2.56) \\ \addlinespace
$^{28}$SiO $v$ = 1, $J$ = 3--2 &  1.96\xx    & 0.05 &  4.34\ee    & 10.7\ee     & $-$0.3\nn\xx  & 120527(2.56) \\
                               &  7.17\xx    & 0.01 & 24.98\ee    & 11.4\ee     &    0.4\nn\xx  & 130410(3.13) \\
                               &  3.55\xx    & 0.02 & 10.38\ee    & 10.7\ee     & $-$0.3\nn\xx  & 130915(3.42) \\
                               &  4.02\xx    & 0.02 & 11.26\ee    & 10.1\ee     & $-$0.9\nn\xx  & 131125(3.54) \\
                               &  0.99\xx    & 0.02 &  2.48\ee    & 10.3\ee     & $-$0.7\nn\xx  & 140211(3.68) \\
                               &  3.32\xx    & 0.01 & 11.56\ee    & 10.1\ee     & $-$0.9\nn\xx  & 140407(3.78) \\
                               &  2.83\xx    & 0.03 & 10.55\ee    & 10.1\ee     & $-$0.9\nn\xx  & 140604(3.89) \\ \addlinespace
$^{28}$SiO $v$ = 2, $J$ = 3--2 & $\cdots$\ee & 0.07 & $\cdots$\nn & $\cdots$\nn & $\cdots$\nn\e & 120527(2.56) \\ \addlinespace
$^{29}$SiO $v$ = 0, $J$ = 1--0 &  0.53\xx    & 0.02 &  1.57\ee    & 10.7\ee     & $-$0.3\nn\xx  & 101211(1.60) \\
                               &  1.02\xx    & 0.05 &  2.37\ee    & 10.9\ee     & $-$0.1\nn\xx  & 110222(1.74) \\
                               &  2.01\xx    & 0.02 &  3.11\ee    & 11.4\ee     &    0.4\nn\xx  & 111018(2.16) \\
                               &  1.50\xx    & 0.02 &  2.44\ee    & 11.4\ee     &    0.4\nn\xx  & 120102(2.30) \\
                               &  0.51\xx    & 0.03 &  0.70\ee    & 11.4\ee     &    0.4\nn\xx  & 120413(2.48) \\

\bottomrule
\end{tabular}
\end{table*}

%%%%%%%%%%%%%%%%%%%%%%%%%%%%%%%%%%%%%%%%%%%%%%%%%%%%%%%%%%%%%%%%%%%%%%%%%%%%%%%%%%%%%%%%%%%%%%%%%%%%%%%%%

%%% FIGURE 1 %%%%%%%%%%%%%%%%%%%%%%%%%%%%%%%%%%%%%%%%%%%%%%%%%%%%%%%%%%%%%%%%%%%%%%%%%%%%%%%%%%%%%%%%%%%%
\begin{figure}[t!]
\centering
\includegraphics[width=83mm]{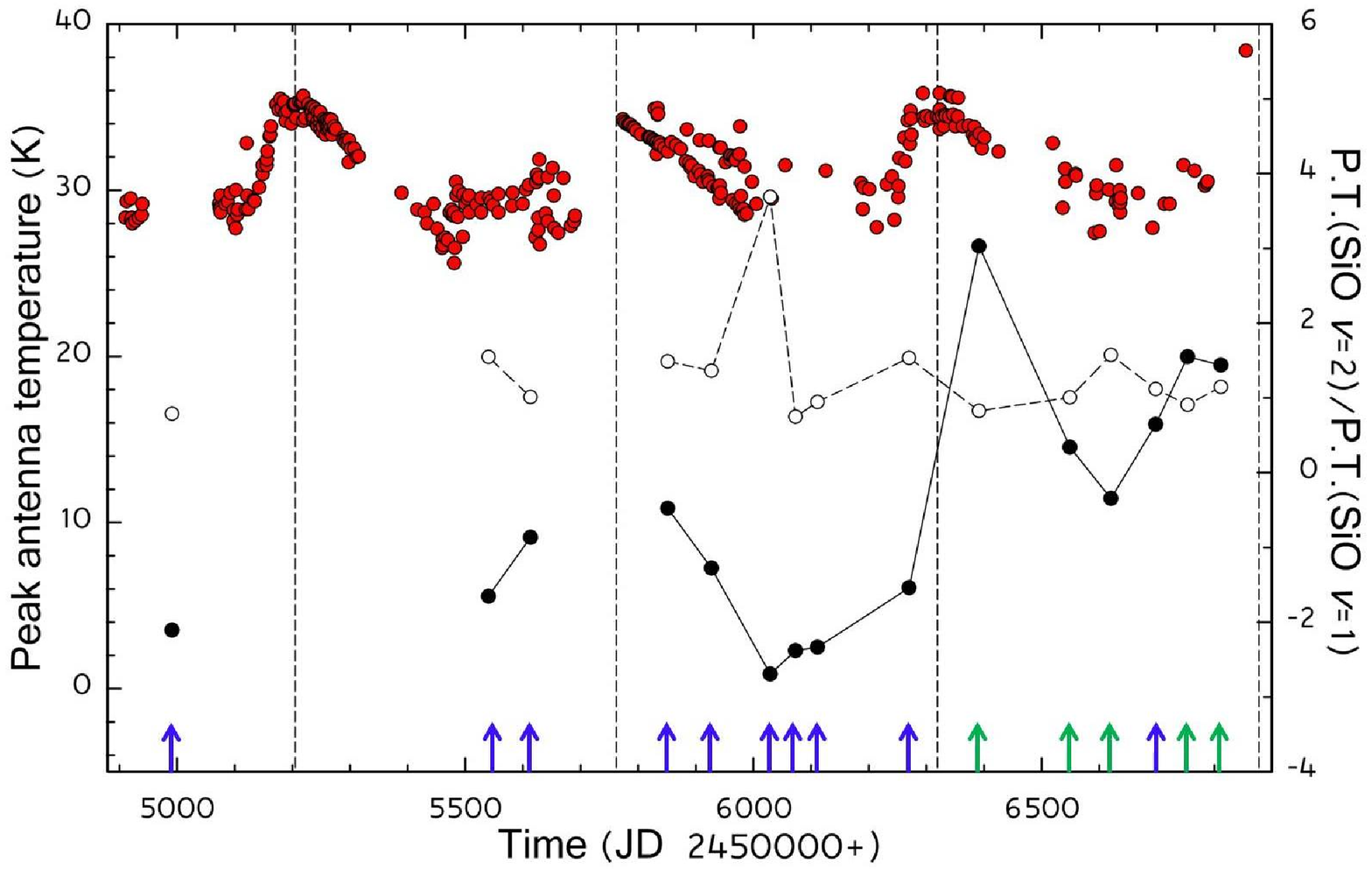}
\caption{Optical light curve of TX Cam from AAVSO (minimum magnitude: $\sim$16.6, maximum magnitude:$\sim$10.4). Optical maxima are indicated by dotted vertical lines. The arrow marks denote our monitoring epochs. The 
H$_{2}$O detected epochs are shown in green. The peak antenna temperature of the SiO $v$ = 1, $J$ = 1--0 maser is plotted according to Julian dates. The peak antenna temperature ratios between the vibrational state 
SiO $v$ = 1 and $v$ = 2, $J$ = 1--0 masers (open circles) are also plotted.\label{fig:jkasfig1}}
\end{figure}
%%%%%%%%%%%%%%%%%%%%%%%%%%%%%%%%%%%%%%%%%%%%%%%%%%%%%%%%%%%%%%%%%%%%%%%%%%%%%%%%%%%%%%%%%%%%%%%%%%%%%%% 

Table 2 presents the results of the SiO and H$_{2}$O monitoring observations. The observed molecules and transitions are given in Column 1. The detected peak antenna temperatures of H$_{2}$O and SiO maser lines and their 
root mean square (rms) levels are given in Columns 2 and 3. Columns 4 and 5 provide integrated antenna temperatures and peak velocities. Column 6 gives the peak velocities with respect to the local standard of rest (LSR) for each detected maser 
line. The dates of the observation with the corresponding phase of the optical light curve (maximum light = 1.0, 2.0, 3.0) are listed in Column 7. The optical phase was calculated from the optical data provided by the American 
Association of Variable Star Observers (AAVSO). 

Figure 1 shows our monitoring epochs with respect to the optical light curve of TX Cam. The H$_{2}$O maser detected epochs are indicated with green arrows. Figure 2 presents all the 
observed spectra of the H$_{2}$O masers according to the observational dates including the detected H$_{2}$O maser lines. As shown in Figure 2 and Table 2, we can confirm the detection of H$_{2}$O maser lines at 5 epochs from April 2013 to 
June 2014. The intensities of the H$_{2}$O maser lines are very weak, from 0.06 K to 0.15 K, and they are detected around the stellar velocity of 11.0 km s$^{-1}$ (Feast \& Whitelock 2000). Figure 3 displays the spectra of H$_{2}$O and SiO 
masers which are simultaneously obtained at the same epoch. 

We confirm that the H$_{2}$O maser emission is very weak compared to SiO masers and also occurs around the peak velocity of SiO masers. These H$_{2}$O maser lines are 
detected from the Mira variable star TX Cam for the first time. The previously undetected H$_{2}$O maser was considered to be related with a transition object from oxygen-rich to carbon-rich stars (Cho \& Ukita 1995). The detection 
of the H$_{2}$O maser in relation with the characteristics of TX Cam will be discussed in the next section.

In Figure 3, one can compare line profile features among SiO $J$ = 1--0, $J$ = 2--1, and $J$ = 3--2 masers. The spectra of SiO $J$ = 1--0, $J$ = 2--1, and $J$ = 3--2 masers exhibit different intensities and peak velocities 
even at same epoch according to their different maser conditions. In addition, SiO $v$ = 1, 2, $J$ = 1--0 spectra were not repeated at similar optical phases as shown in spectra at the epochs June 10, 2009 (090610, $\phi$ = 0.62), 
December 11, 2011 (111211, $\phi$ = 1.60), July 4, 2012 (120704, $\phi$ = 2.63), and February 11, 2014 (140211, $\phi$ = 3.68). The SiO $v$ = 0, $J$ = 1--0, $J$ = 2--1 spectra relative to May 27, 2012 (120527) and July 4, 2012 (120704) show a 
broad and parabolic shape (the FWZP of SiO $v$ = 0, $J$ = 1--0 lines = 37.41, 47.41 km s$^{-1}$, FWZP of $v$ = 0, $J$ = 2--1 line = 47.5 km s$^{-1}$) compared to the SiO $v$ = 1, 2, $J$ = 1--0, $J$ = 2--1, and $J$ = 3--2 maser 
lines (the FWZP of $v$ = 1, 2, $J$ = 1--0 masers = 23.89, 22.75 km s$^{-1}$, and FWZP of $v$ = 1, $J$ = 2--1 maser = 18.66 km s$^{-1}$) because they originate from thermal emission. In addition, the 
SiO $v$ = 0, $J$ = 1--0 spectra show a combination of thermal parabolic and spike emission due to partial maser emission. The expansion velocity of TX Cam can be measured to be $\sim$ 23.7 km s$^{-1}$ from the half of 
the full width at zero power of these thermal lines.
The peak antenna temperature variations of SiO $v$ = 1, $J$ = 1--0 masers are shown in Figure 1, together with the optical light curve. The behavior of the SiO intensity curves corresponds well with the optical light curve, although 
the monitoring time interval and numbers are limited. However, it is difficult to confirm the phase lag of SiO maxima with respect to the optical maxima although the SiO intensity shows a maximum on April 10, 2013 (at $\phi$ = 3.13, 
the first detection phase of H$_{2}$O) because our monitoring epochs around the optical maxima are very limited. The SiO $v$ = 2, $J$ = 2--1 and $J$ = 3--2 masers were not detected near the minimum optical phase 
$\phi$ = 2.56 (May 27, 2012) as shown in Figure 3.

\section{Discussion\label{sec:disc}}

\subsection{Detection of the H$_{2}$O Maser Emission\label{sec:detect}}

%%% FIGURE 2 %%%%%%%%%%%%%%%%%%%%%%%%%%%%%%%%%%%%%%%%%%%%%%%%%%%%%%%%%%%%%%%%%%%%%%%%%%%%%%%%%%%%%%%%%%%%
\begin{figure*}[t!]
\centering
\includegraphics[width=170mm]{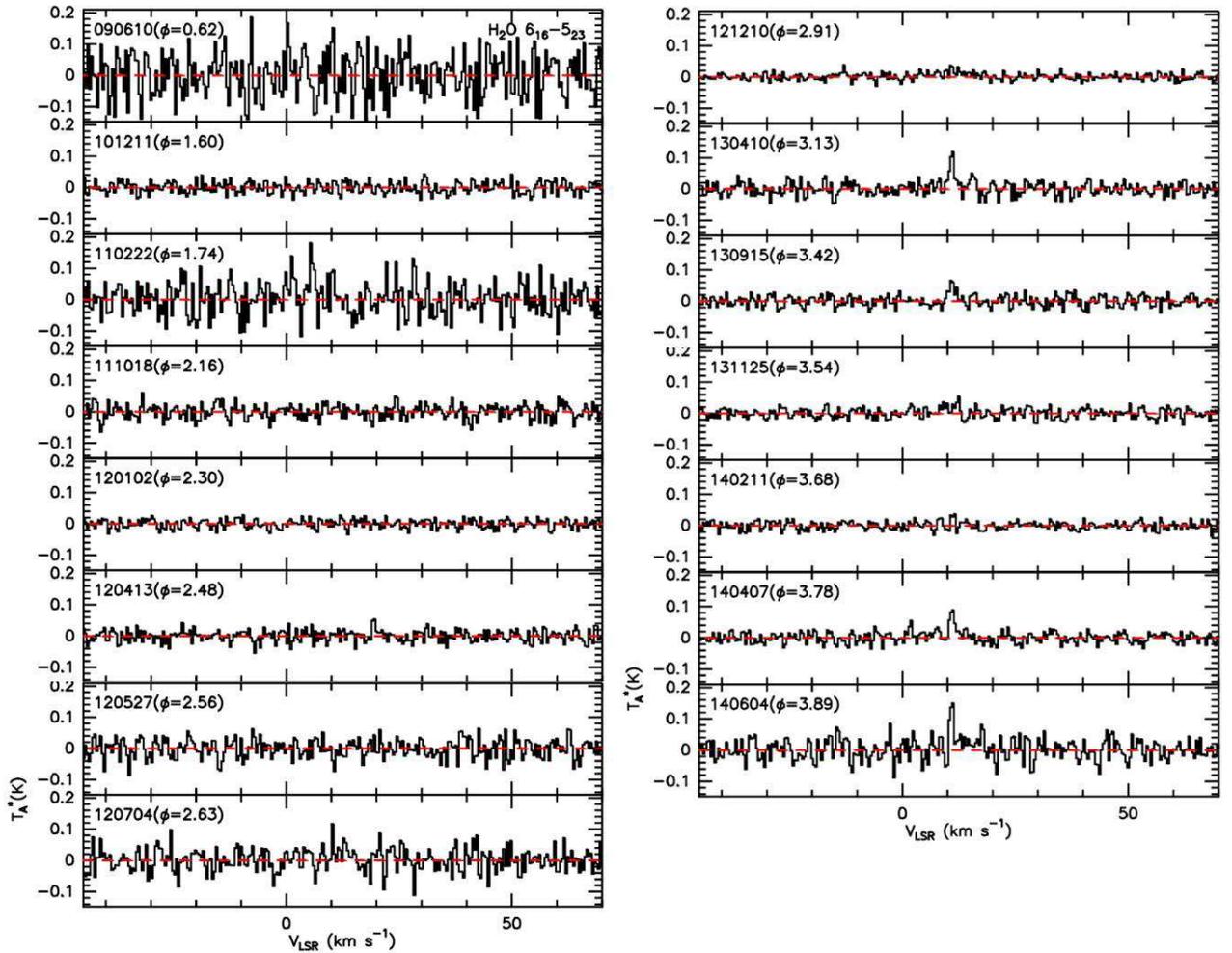}
\caption{The totality of H$_{2}$O maser spectra observed from June 10, 2009 to June 4, 2014 (15 epochs). The intensity is given in units of the antenna temperature $T_{\rm A}^{\ast}$ (K) and the abscissa is $V_{\rm LSR}$ (km s$^{-1}$). 
The date of observation and optical phase are also given in the spectrum of each source. The H$_{2}$O maser was detected at five epochs ($\phi$ = 3.13, 3.42, 3.54, 3.78, 3.89). \label{fig:jkasfig2}}
\end{figure*}
%%%%%%%%%%%%%%%%%%%%%%%%%%%%%%%%%%%%%%%%%%%%%%%%%%%%%%%%%%%%%%%%%%%%%%%%%%%%%%%%%%%%%%%%%%%%%%%%%%%%%%%

%%% FIGURE 3 %%%%%%%%%%%%%%%%%%%%%%%%%%%%%%%%%%%%%%%%%%%%%%%%%%%%%%%%%%%%%%%%%%%%%%%%%%%%%%%%%%%%%%%%%%%%
\begin{figure*}[tp]
\centering
\includegraphics[width=160mm]{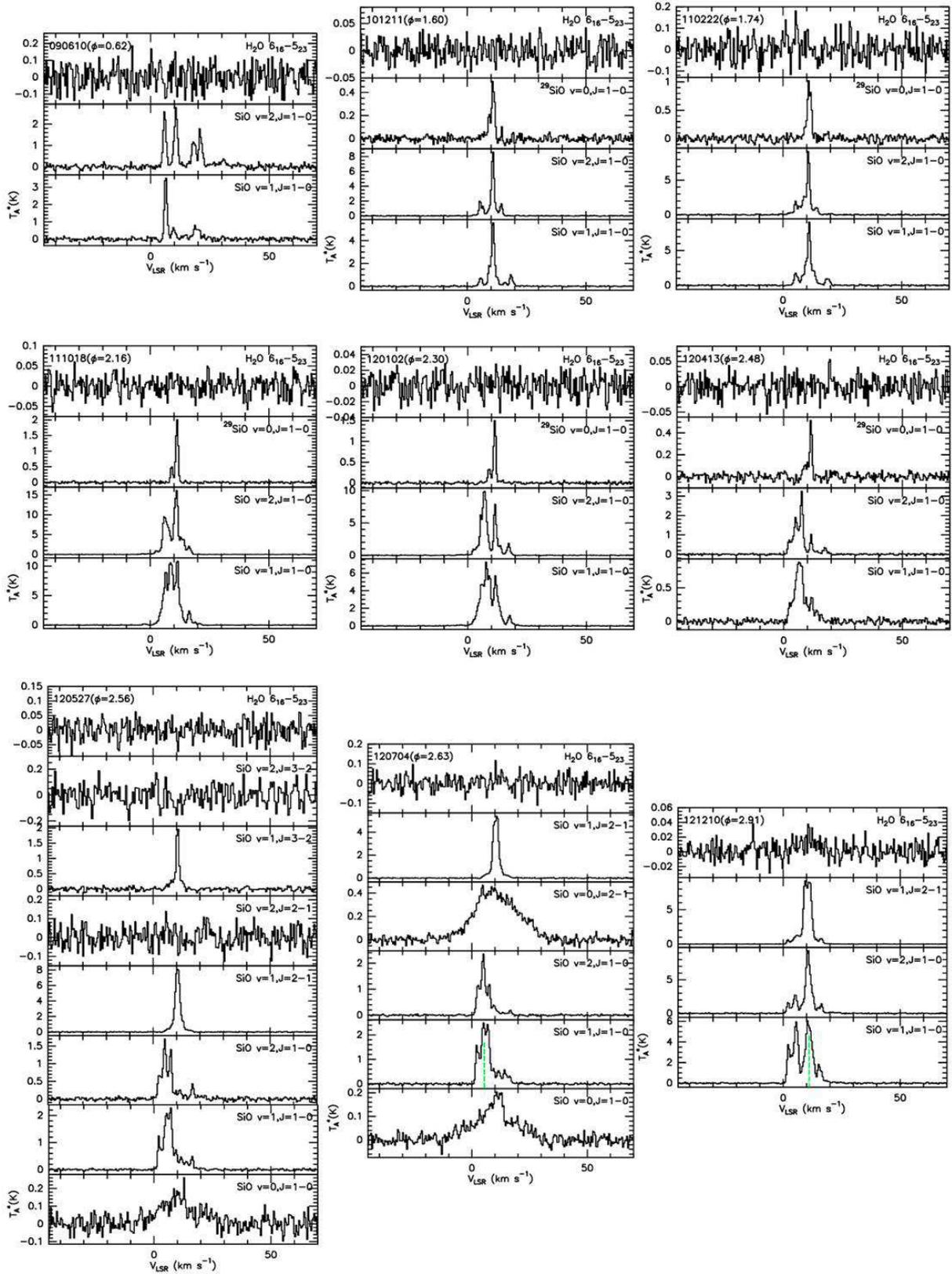}
\caption{The H$_{2}$O and SiO spectra simultaneously obtained in the H$_{2}$O $6_{16}-5_{23}$ and SiO $v$ =0, 1, 2, $J$ = 1--0, $J$ = 2--1, $J$ = 3--2, and $^{29}$SiO $v$ = 0, $J$ = 1--0 lines during June 10, 
2009 -- June 4, 2014. The spectra are arranged in the same order as in the Table 2. Intensity is given in units of antenna temperature $T_{\rm A}^{\ast}$ (K) and the abscissa is $V_{\rm LSR}$ (km s$^{-1}$). The strongest peaks are indicated with green dotted lines in the spectra. \label{fig:jkasfig3}}
\end{figure*}

\setcounter{figure}{2}
\begin{figure*}[t!]
\centering
\includegraphics[width=160mm]{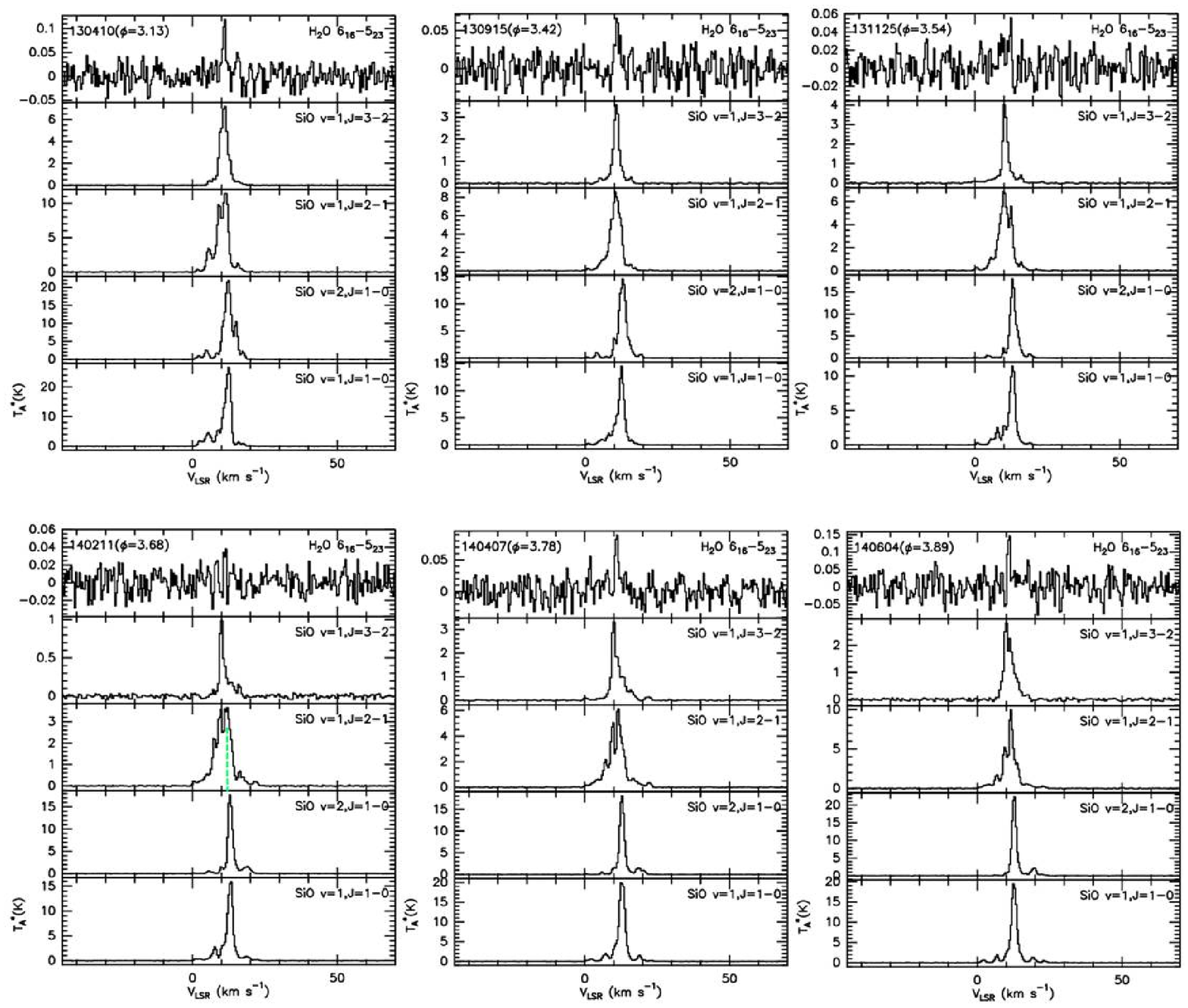}
\caption{(Continued)\label{fig:jkasfig3-cont}}
\end{figure*}
%%%%%%%%%%%%%%%%%%%%%%%%%%%%%%%%%%%%%%%%%%%%%%%%%%%%%%%%%%%%%%%%%%%%%%%%%%%%%%%%%%%%%%%%%%%%%%%%%%%%%%%

The H$_{2}$O maser emission has started to be detected from April 10, 2013 ($\phi$ = 3.13). Our high sensitivity observations (0.4--1.2 Jy in 2013 -- 2014) together with our 
monitoring observations over three periods extended from the optical maximum to minimum phases led us to detect the H$_{2}$O maser. The detection limits of the Benson \& Little-Marenin (1996) and Shintani et al. (2008) 
observations were about or above 1 Jy. Another possibility for the detection of the H$_{2}$O maser may be a change of physical and chemical conditions for pumping the H$_{2}$O maser in TX Cam since 2013. However, it is 
difficult to accept those rapid changes without confirming previous long-term monitoring results of H$_{2}$O masers and other wavelength spectroscopic results. TX Cam showed unusual infrared and chemical characteristics 
for a known oxygen-rich AGB star although it has shown a normal feature as a Mira variable star. These unusual characteristics were discussed by Cho \& Ukita (1995), and this is why TX Cam is located in the region VII of the IRAS two-color 
diagram (van der Veen \& Habing 1988) due to a large IRAS flux ratio of S(60)/S(25). The region VII contains a large number of variable stars with C-rich circumstellar shells. In the case of TX Cam, carbon-bearing molecules 
are detected with particularly high abundances (Lindqvist et al. 1988, 1992). Therefore, this star is considered as a transition object between oxygen-rich and carbon-rich stars. If this star is a transition object, 
for example, in a similar stage as an S star, the H$_{2}$O maser emission may not be detected, as well as in a typical S star $\chi$ Cyg. However, the H$_{2}$O maser was detected despite its intensity (peak antenna 
temperature: 0.06 -- 0.15 K) compared with other Mira variable stars. In particular, the intensity ratios of H$_{2}$O with respect to SiO ($v$ = 1, $J$ = 1--0) masers are very low, ranging from 0.0045 to 0.0077. These intensity 
ratios are above 0.1 for almost all of AGB stars (Kim et al. 2010). Exceptions appear in R Cas and R Leo, which also have very low intensity ratios less than 0.1 at most optical phases. 

The Mira variable stars TX Cam, R Cas, and R Leo show the rare maser SiO $v$ = 2, $J$ = 2--1 line. Olofsson et al. (1981) explained that the weakness of the $v$ = 2, $J$ = 2--1 line in oxygen-rich stars may be attributed 
to an infrared line overlap of H$_{2}$O with the $v$ = 1, $J$ = 0 $\rightarrow$ $v$ = 2, $J$ = 1 line. Therefore, they suggested that for these H$_{2}$O deficient stars, the rare maser SiO $v$ = 2, $J$ = 2--1 line can occur 
without the line overlap disturbance. The SiO $v$ = 2, $J$ = 2--1 maser emission is also detected from the S star $\chi$ Cyg in which the H$_{2}$O maser is not detected. Cho \& Ukita (1995) also suggested that the detection of 
another rare maser $^{29}$SiO $v$ = 1, $J$ = 1--0 line from TX Cam can be explained by a similar line overlap of H$_{2}$O. In these stars, enough oxygen may not be available for forming H$_{2}$O molecule as suggested 
by Benson \& Little-Marenin (1996). In addition, long-term and regular monitoring observations of 22 GHz H$_{2}$O maser are required for tracing the continuous detectability and the variations of intensities according to the optical 
phases together with sub-mm H$_{2}$O maser observations.

\subsection{Intensity Ratio and Peak Velocity Pattern among SiO Masers According to the Optical Phases\label{sec:ratio}}

The peak antenna temperature ratios between the vibrational state SiO $v$ = 1 and $v$ = 2, $J$ = 1--0 masers are plotted in Figure 1. The variation of $v$ = 2 / $v$ = 1 peak temperature ratio does not show a large spread with 
respect to the ratio 1 except the ratio 3.69 on April 13, 2012 ($\phi$ = 2.48). The peak antenna temperature of the SiO $v$ = 1, $J$ = 1--0 maser at $\phi$ = 2.48 is quite low. The average value of these ratios at 14 epochs 
(except the value at $\phi$ = 2.48) is about 1.14. Furthermore, the peak antenna temperatures of SiO $v$ = 2, $J$ = 1--0 show large values at 10 epochs among 15 observed epochs compared to those of $v$ = 1. 
Alcolea et al. (1999) reported that the $v$ = 2 / $v$ = 1 integrated flux ratio is almost unity independently of the phase. This intensity ratio may be associated with the evolutionary stage of a late AGB phase as indicated by 
Cho \& Kim (2012). The intensity of the SiO $v$ = 2 maser can increase according to the development of a hot dust shell from a late AGB phase to a post-AGB one (Cho \& Kim 2012). This tendency clearly appeared in post-AGB stars 
accompanying only $v$ = 2, $J$ = 1--0 maser detections without $v$ = 1 detections (Yoon et al. 2014). Therefore, this tendency seems to appear in the very late M-type star TX Cam (M8--M10).

%%% TABLE 3 %%%%%%%%%%%%%%%%%%%%%%%%%%%%%%%%%%%%%%%%%%%%%%%%%%%%%%%%%%%%%%%%%%%%%%%%%%%%%%%%%%%%%%%%%%%%
\begin{table*}[t!]
\caption{Variation of peak antenna temperature ratios of SiO $v$ = 1 masers according to their optical phases\label{tab:jkastable3}}
\centering
\begin{tabular}{cccc}
\toprule
Date(phase) & \large $\frac{\rm P.T.(SiO,J=2-1)}{\rm P.T.(SiO,J=1-0)}$ & \large $\frac{\rm P.T.(SiO,J=3-2)}{\rm P.T.(SiO,J=1-0)}$ & \large $\frac{\rm P.T.(SiO,J=3-2)}{\rm P.T.(SiO,J=2-1)}$ \\
\midrule
120527(2.56) & 3.56 & 0.86     & 0.24     \\
120704(2.63) & 2.17 & $\cdots$ & $\cdots$ \\
121210(2.91) & 1.51 & $\cdots$ & $\cdots$ \\
130410(3.13) & 0.43 & 0.27     & 0.63     \\
130915(3.42) & 0.59 & 0.24     & 0.41     \\
131125(3.54) & 0.60 & 0.35     & 0.59     \\
140211(3.68) & 0.23 & 0.06     & 0.27     \\
140407(3.78) & 0.31 & 0.17     & 0.54     \\
140604(3.89) & 0.51 & 0.15     & 0.28     \\
\bottomrule
\end{tabular}
\end{table*}
%%%%%%%%%%%%%%%%%%%%%%%%%%%%%%%%%%%%%%%%%%%%%%%%%%%%%%%%%%%%%%%%%%%%%%%%%%%%%%%%%%%%%%%%%%%%%%%%%%%%%%%%%

In Table 3, we provide the variation of the peak antenna temperature ratios among SiO $v$ = 1, $J$ = 1--0, $J$ = 2--1, and $J$ = 3--2 masers, according to nine optical phases of our monitoring observations. The peak antenna 
temperature ratios of $J$ = 2--1 / $J$ = 1--0 exhibit a large spread compared to those of P.T.($J$ = 3--2) / P.T.($J$ = 1--0) and P.T.($J$ = 3--2) / P.T.($J$ = 2--1). In particular, those peak ratios 
exhibit a large spread from the second to the third period, despite similar optical phases. These differences seem to be related with a SiO maser pumping mechanism, which responds to 
different local physical conditions among different SiO maser lines. The average peak antenna temperature ratios among SiO $v$ = 1, $J$ = 1--0, $J$ = 2--1, and $J$ = 3--2 masers are estimated to be 
P.T.($J$ = 2--1) / P.T.($J$ = 1--0) = $\sim$ 0.85, P.T.($J$ = 3--2) / P.T.($J$ = 1--0) = $\sim$ 0.30, and P.T.($J$ = 3--2) / P.T.($J$ = 2--1) = $\sim$ 0.42, respectively. These SiO intensity ratios can be deeply 
influenced by dust, as shown in the Gray et al. (2009) model. The previous authors suggested that collisional and radiative pumping are spatially associated, via close association of shock and radii of the SiO pumping band 
at 8.13 micron of dust. The peak antenna temperature ratios obtained from simultaneous monitoring observations of multi-frequencies will provide good constraints for these maser pumping models.

In Figure 4, the shift of the peak velocities of the 22 GHz H$_{2}$O and SiO masers with respect to the stellar velocity is shown according to the observed optical phases. As clear from the figure, the H$_{2}$O maser occurs near the stellar velocity during our monitoring interval. On the other hand, SiO masers show a spread in both the blue- and red-shifted part with respect to the stellar velocity. One of the reasons can be the difference of H$_{2}$O and SiO maser regions. Namely, SiO masers arise from a close to central star 2--4 $R_{\odot}$ (Diamond et al. 1994) which is directly influenced by pulsating motion. The H$_{2}$O masers occur 
farther out from above dust formation layers where the expansion velocity approaches the terminal velocity of the mass loss. 

Among SiO masers, the peak velocities of SiO $J$ = 2--1, and $J$ = 3--2 masers show a smaller spread with respect to the stellar velocity than those of SiO $J$ = 1--0 masers. It is difficult to discuss the systematic pulsation 
effect on the shift of SiO peak velocities due to limited monitoring data. Simultaneous monitoring observations of SiO $J$ = 2--1 and $J$ = 3--2 masers using the TRAO 14 m telescope (Kang et al. 2006) showed that the main peak 
component of the $v$ = 1, $J$ = 2--1 maser has dominant red-shifted features compared to the $v$ = 1, $J$ = 3--2 maser. These differences may originate from the differences of maser regions among SiO $J$ = 1--0, $J$ = 2--1, 
and $J$ = 3--2 masers including the differences of masing conditions. The VLBA observations of SiO $v$ = 1, $J$ = 1--0 and $J$ = 2--1 masers showed that the SiO $v$ = 1, $J$ = 2--1 arise farther away than the SiO $v$ = 1, 
$J$ = 1--0 in the case of the OH/IR star WX Psc (Soria-Ruiz et al. 2004). However, they arise from almost the same distance from the central star for $\chi$ Cyg. These VLBI observations of SiO $J$ = 1--0 and $J$ = 2--1 masers are very limited due to lack of an existing VLBI system which can be operated at both SiO $J$ = 1--0 and $J$ = 2--1 maser observing bands. Furthermore, except for the KVN, there are no VLBI systems in the world that can be 
operated in the SiO $J$ = 3--2 observing band. Therefore, KVN will play an important role in near future.  

%%% FIGURE 4 %%%%%%%%%%%%%%%%%%%%%%%%%%%%%%%%%%%%%%%%%%%%%%%%%%%%%%%%%%%%%%%%%%%%%%%%%%%%%%%%%%%%%%%%%%%%
\begin{figure}[t!]
\centering
\includegraphics[width=83mm]{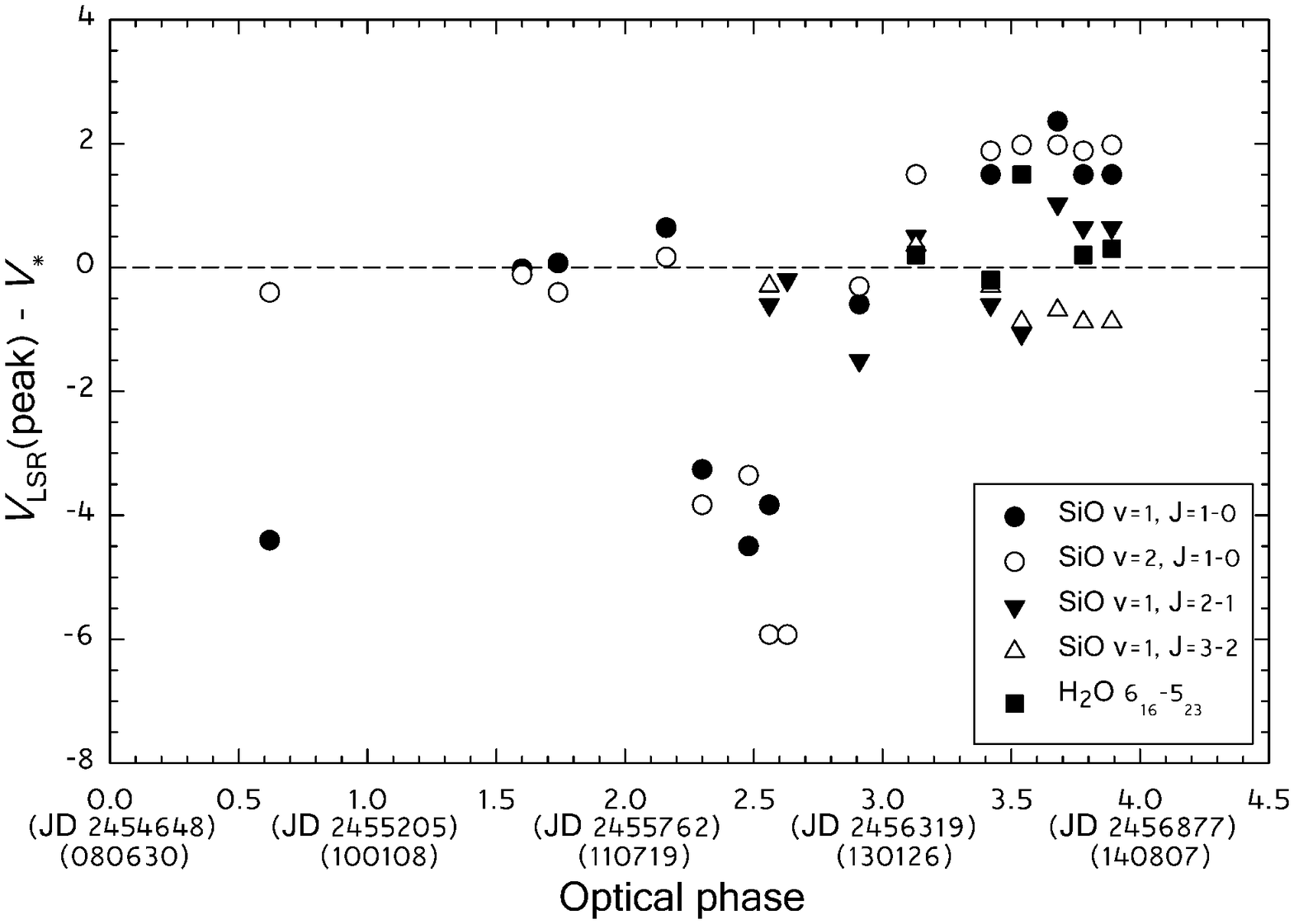}
\caption{Shift of peak velocities of 22 GHz H$_{2}$O and SiO masers with respect to the stellar velocity, according to optical phases (Julian dates and observation dates). Each maser line is indicated by different symbols 
in the square box. The stellar velocity $V_{\ast}$ is 11.0 km s$^{-1}$. \label{fig:jkasfig4}}
\end{figure}
%%%%%%%%%%%%%%%%%%%%%%%%%%%%%%%%%%%%%%%%%%%%%%%%%%%%%%%%%%%%%%%%%%%%%%%%%%%%%%%%%%%%%%%%%%%%%%%%%%%%%%%

\section{Summary\label{sec:sum}}

\e Our results can be summarized as follows. 

1. We detected the 22 GHz H$_{2}$O maser emission from a Mira variable star TX Cam for the first time. The 22 GHz H$_{2}$O maser was detected near the stellar velocity at five epochs from April 10, 2013 ($\phi$ = 3.13) 
to June 4, 2014 ($\phi$ = 3.89) including the period around minimum optical phases. The peak antenna temperatures were very weak (0.06 -- 0.15 K) compared to SiO $v$ = 1, $J$ = 1--0 masers (the peak antenna temperature ratio between 
P.T.(SiO $v$ = 1, $J$ = 1--0) / P.T.(H$_{2}$O) = 0.0045 at minimum). These weak intensities of the H$_{2}$O maser emission may suggest that the existence of rare masers SiO $v$ = 2, $J$ = 2--1 and $^{29}$SiO $v$ = 1, 
$J$ = 1--0 lines is associated with an infrared line overlap of H$_{2}$O. 

2. The average value of peak antenna temperature ratios between the vibrational state SiO $v$ = 1 and $v$ = 2, $J$ = 1--0 masers at 14 epochs (except the value at $\phi$ = 2.48) is about 1.14. Furthermore, the peak antenna 
temperatures of SiO $v$ = 2, $J$ = 1--0 show large values at 10 epochs among 15 observed epochs compared to those of $v$ = 1. The intensity of SiO $v$ = 2 maser can increase according to the development of a hot dust shell from 
a late AGB phase to a post-AGB one. Therefore, this tendency seems to appear in the very late M-type star TX Cam (M8--M10).

3. The peak antenna temperature ratios of $J$ = 2--1 / $J$ = 1--0 show a large spread compared to those of P.T.($J$ = 3--2) / P.T.($J$ = 1--0) and P.T.($J$ = 3--2) / P.T.($J$ = 2--1). In particular, the peak antenna temperature 
ratios of $J$ = 2--1 / $J$ = 1--0 exhibit a large spread despite similar optical phases in the second and third periods. The average peak antenna temperature ratios among SiO $v$ = 1, $J$ = 1--0, $J$ = 2--1, and $J$ = 3--2 masers 
are estimated to be P.T.($J$ = 2--1) / P.T.($J$ = 1--0) = $\sim$ 0.85, P.T.($J$ = 3--2) / P.T.($J$ = 1--0) = $\sim$ 0.30, and P.T.($J$ = 3--2) / P.T.($J$ = 2--1) = $\sim$ 0.42, respectively. 

4. The peak velocities of SiO $J$ = 2--1 and $J$ = 3--2 masers occur closer to the stellar velocity than those of SiO $J$ = 1--0 masers. These differences may be originated from the differences of maser regions among 
SiO $J$ = 1--0, $J$ = 2--1, and $J$ = 3--2 masers including the differences of masing conditions. The simultaneous VLBI monitoring observations of SiO $J$ = 1--0, $J$ = 2--1, and $J$ = 3--2 masers are highly required 
for determining the relative distributions and the variation of these masers, according to the stellar pulsation. 

\acknowledgments
This work was supported by Basic and Fusion Research Programs (2010--2014) and also partially supported by KASI--Yonsei Joint Research Program ``Degree and Research Center Program'' funded by the National Research Council 
of Science and Technology (NST). In this research we have used the information from the AAVSO International Database operated at AAVSO Headquarter, 25 Birch Street, Cambridge, MA 02138 and SIMBAD database operated at CDS, 
Strasbourg, France.

\end{document}